\documentclass[11pt,twoside]{article}
\usepackage{asp2010}

\resetcounters

\begin{document}

\title{Observational study of the extremely slow nova V1280 Scorpii}
\author{H. Naito,$^1$ S. Mizoguchi,$^2$ A. Arai,$^3$ A. Tajitsu,$^4$ S. Narusawa,$^3$ M. Yamanaka,$^5$ M. Fujii,$^6$ T. Iijima,$^7$ K. Kinugasa,$^8$ M. Kurita,$^9$ T. Nagayama,$^1$ H. Yamaoka,$^{10}$ and K. Sadakane$^{11}$
\affil{$^1$Graduate School of Science, Nagoya University, Furo-cho, Chikusa-ku, Nagoya 464-8602, Japan}
\affil{$^2$Sendai Astronomical Observatory, Nishikigaoka, Aoba-ku, Sendai 989-3123, Japan}
\affil{$^3$Nishi-Harima Astronomical Observatory, University of Hyogo, Sayo-cho, Hyogo 679-5313, Japan}
\affil{$^4$Subaru Telescope, National Astronomical Observatory of Japan, 650 North A'ohoku Place, Hilo, HI 96720, USA}
\affil{$^5$Kwasan Observatory, Kyoto University, Kitakazan-ohmine-cho, Yamashina-ku, Kyoto 607-8471, Japan}
\affil{$^6$Fujii Kurosaki Observatory, 4500 Kurosaki, Tamashima, Kurashiki, Okayama 713-8126, Japan}
\affil{$^7$Astronomical Observatory of Padova, Asiago Section, Osservatorio Astrofisico, 36012 Asiago (Vi), Italy}
\affil{$^8$Nobeyama Radio Observatory, National Astronomical Observatory of Japan, Minamimaki, Minamisaku-gun, Nagano 384-1305, Japan}
\affil{$^9$Graduate School of Science, Kyoto University, Kitashirakawaoiwake-cho, Sakyo-ku, Kyoto 606-8502, Japan}
\affil{$^{10}$Graduate School of Sciences, Kyushu University, Hakozaki, Higashi-ku, Fukuoka 812-8581, Japan}
\affil{$^{11}$Astronomical Institute, Osaka Kyoiku University, Asahigaoka, Kashiwara, Osaka 582-8582, Japan}}

\begin{abstract}
We present multi-color light curves and optical spectra of V1280 Scorpii obtained from 2007 to 2012. It is shown that V1280 Sco is the extremely slow nova and the mass of white dwarf appears to be $\sim$ 0.6 M$\odot$ or lower. Blue-shifted multiple absorption lines of Na {\sc i} D, Ca  {\sc ii}  HK, and He  {\sc i*}  are detected on high-resolution spectra. We also discuss that an approach using metastable He absorption lines is useful to investigate structures of nova shells.
\end{abstract}

\section{Introduction}
V1280 Scorpii is a classical nova which was independently discovered by two Japanese amateur astronomers (Y. Nakamura and Y. Sakurai) at the position of R.A. = 16$^\mathrm{h}$57$^\mathrm{m}$41$^\mathrm{s}$.0, Decl. = $-$32$^{\circ}$20' 36".4 (equinox 2000.0)  on 2007 February 4 at ninth visual magnitude \citep{2007IAUC.8803....1Y}. \citet{2007IAUC.8803....2N} obtained a low dispersion spectrum on February 5.87 (one day after the discovery), and found that it had a smooth continuum together with the Balmer and Fe{\sc ii} lines showing P Cygni profiles and confirmed that it was classified as a classical nova. After maximum light, it faded steadily for about 12 days, before undergoing a precipitous decline in visual magnitude caused by dust formation. As the dust shells expand and spread out, it became gradually brighter and it has kept a brightness at $V$ and $y$ magnitudes of each around 10 mag to this date (2012 August). V1280 Sco has a very unique light curve which has not been observed in novae before. We performed early follow-up spectroscopic observations mainly at the Nishi-Harima Astronomical Observatory, long-term photometric observations mainly at Osaka Kyoiku University, and high-dispersion spectroscopic observations at the Subaru Telescope.

\section{Extremely Slow Evolution}
The long-term photometric observations show that V1280 Sco decreased in brightness gradually at a very slow rate from its discovery to our last photometric observation in 2012; and that it has a very long plateau spanning over 1500 days in its light curve (Fig. 1). Furthermore, long-term spectroscopic observations show that the transition to the nebula phase defined by the appearance of both [O  {\sc iii}] 4959 and 5007 took around 50 months after the burst (see Fig. 14 in Naito et al. 2012); and that the wind continued to blow spanning over 1500 days after maximum light. By comparing V1280 Sco with other slow novae reported in the literature (V723 Cas; \citeauthor{2006A&A...451..563I} \citeyear{2006A&A...451..563I}, \citeauthor{2004ApJ...612L..57H} \citeyear{2004ApJ...612L..57H}, GQ Mus; \citeauthor{2008ApJ...687.1236H} \citeyear{2008ApJ...687.1236H}), we conclude that this nova is going through the slowest evolution in history and the mass of white dwarf is likely to be $\sim$ 0.6 M$\odot$ or lower \citep{2012A&A...543A..86N}.

\articlefigure{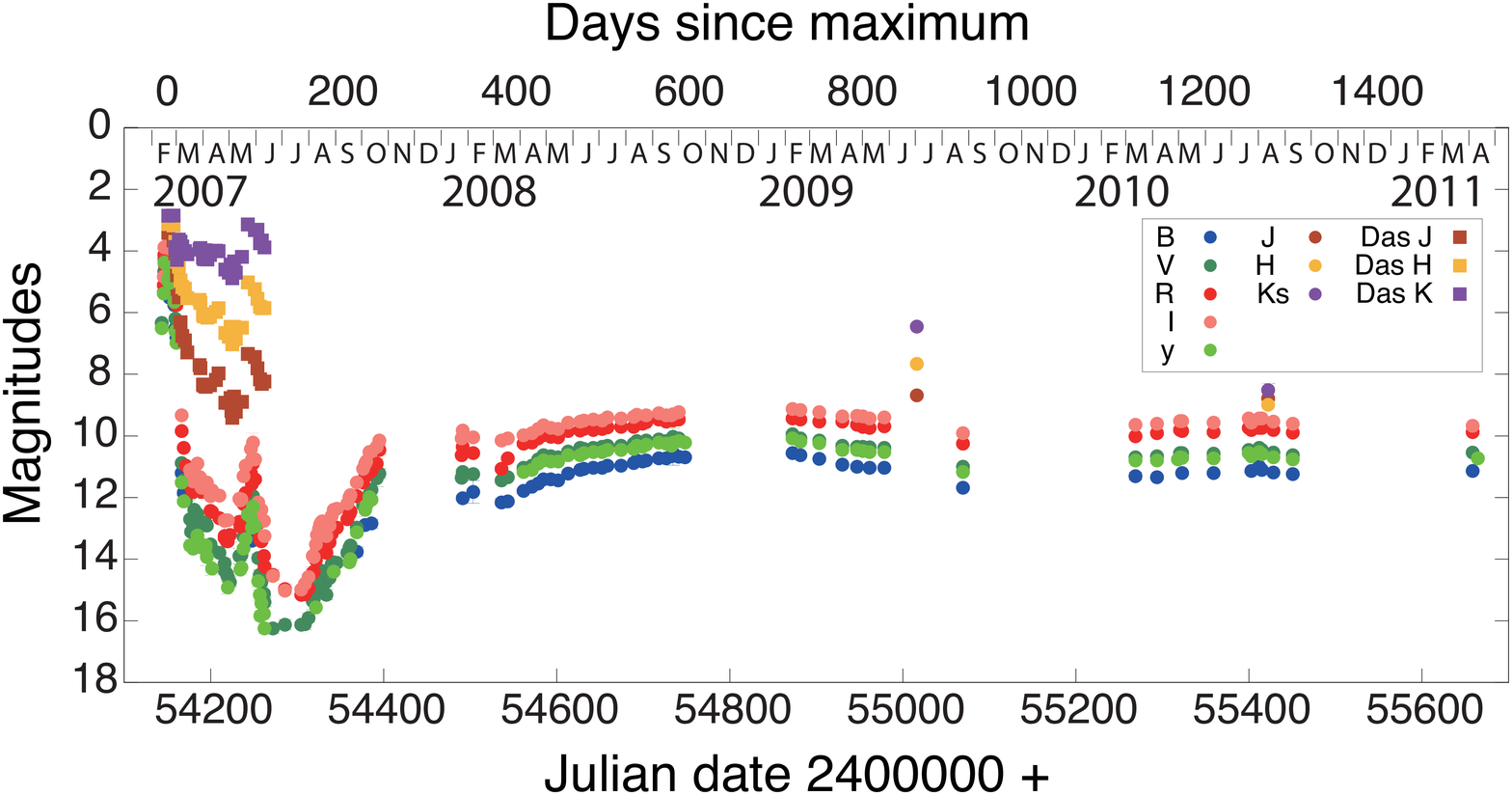}{}{Light curves in optical $B$, $V$, $R_{\rm c}$, $I_{\rm c}$, and $y$ and NIR $J$, $H$, and $K_{\rm s}$ bands observed from 2007 to 2011 \citep{2012A&A...543A..86N}. NIR light curves in the early phase published by \citet{2008MNRAS.391.1874D} are superimposed.}

\begin{figure}
\plotfiddle{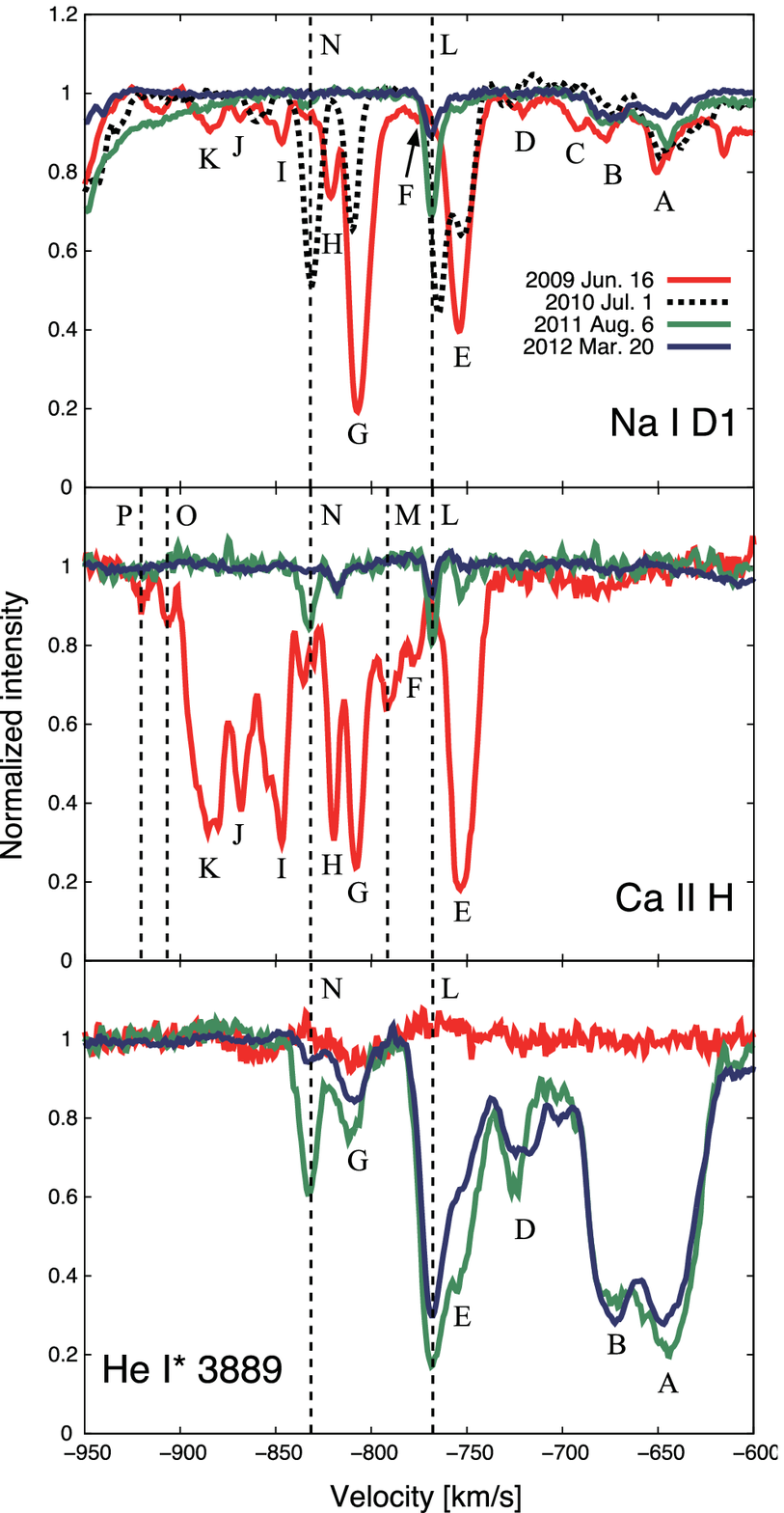}{9.2cm}{0}{75}{53}{-100}{-10}
\caption{Detection of multiple absorption lines of Na {\sc i} D1, Ca {\sc ii} H and He {\sc i*} $\lambda$3889 \citep{2013PASJ...65...37N}.}
\end{figure}

\section{Multiple Absroption Lines}
Multiple blue-shifted metastable He {\sc i*} absorption lines at 3889 $\AA$ are found on the high-resolution spectra (see Fig. 2). This is the first detection of metastable He  {\sc i*}  absorption lines in the ejected gas around novae. We also discovered similar multiple blue-shifted absorption lines of Na D and Ca HK \citep{2010PASJ...62L...5S}. The results of analysis using these lines show that the ejected shells are consist of numerous clumpy gas which cover a significant part of the continuum emitting radiation region; and that the appearance of metastable He  {\sc i*}  lines can be explained by the ionization of helium, caused by the increase in ultraviolet radiation accompanying temperature increase in the photosphere; and that the total mass of the ejected material can be estimated to be on the order of 10$^{-4}$ M$\odot$ \citep{2013PASJ...65...37N}. Analysis using metastable He  {\sc i*}  absorption lines was an unprecedented approach in nova research. We propose that this method could be very useful in measuring helium composition and the mass of shell, and in studying nova shell structures.

\section{Structure of Ejected Shell}
\citet{2012A&A...545A..63C} obtained high-spatial resolution mid-infrared images with the Very Large Telescope from 2009 to 2011, showing the dusty nebular around V1280 Sco is bipolar. It is implied that a large part of ejected mass is included in these bipolar lobes. They suggested that the intensities of two bipolar lobes are different due to an absorption effect; and that the southern lobe is closer to the observer than the northern one.  Combined with the presence of absorbing gases on the line-of-sight shown in our observations, we can describe the structure of ejected shell as Fig. 3 \citep{2013PASJ...65...37N}.

\begin{figure}
\plotfiddle{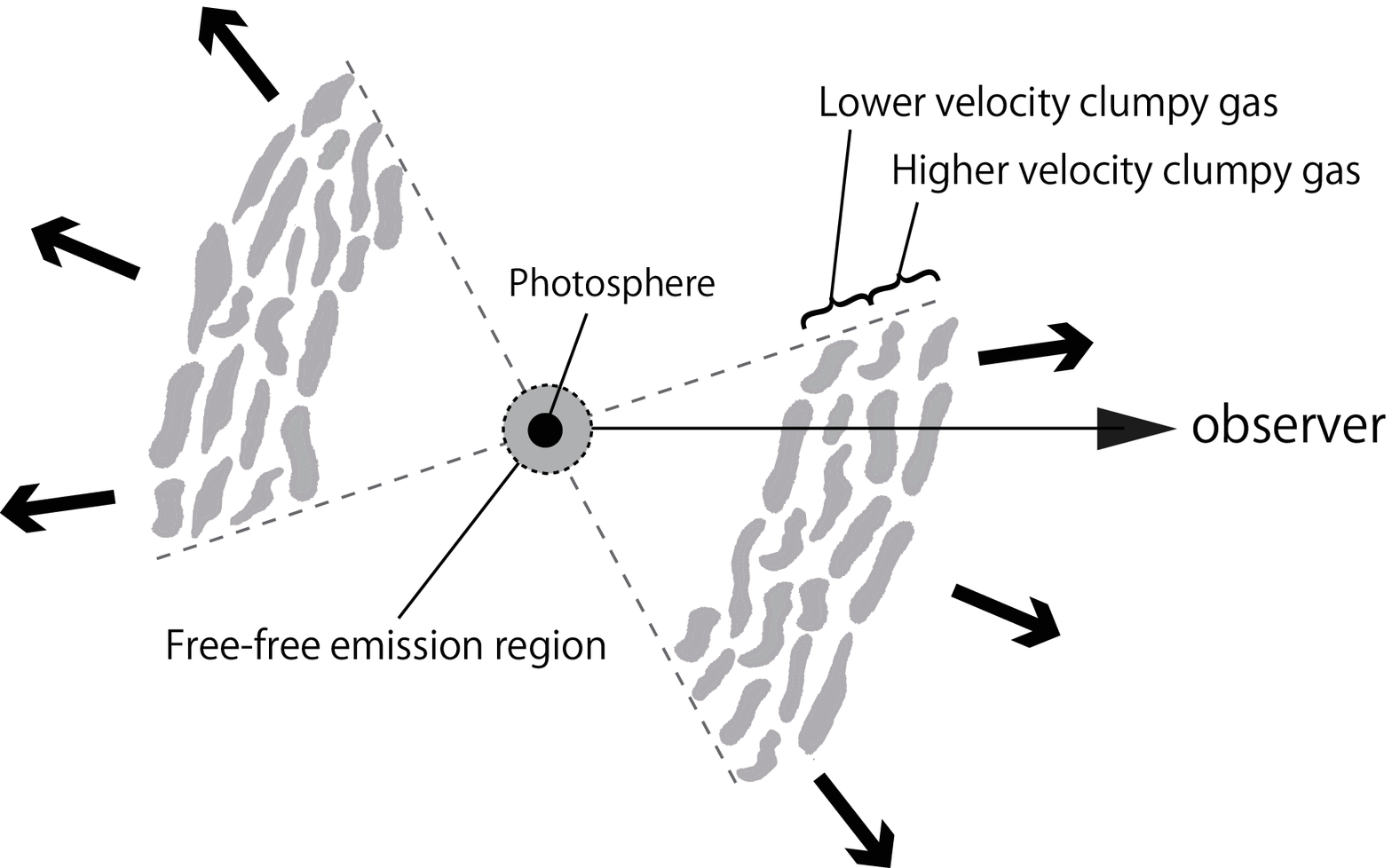}{8cm}{0}{75}{75}{-150}{0}
\caption{A schematic view of ejected shell (clumpy gas) producing absorption lines of metastable He {\sc i*} \citep{2013PASJ...65...37N}. Thin
broken lines delineate boundaries of the nebula \citep{2012A&A...545A..63C}.}
\end{figure}


\bibliography{aspauthor}

\begin{thebibliography}{}
\expandafter\ifx\csname natexlab\endcsname\relax\def\natexlab#1{#1}\fi
\expandafter\ifx\csname url\endcsname\relax
  \def\url#1{\texttt{#1}}\fi
\expandafter\ifx\csname urlprefix\endcsname\relax\def\urlprefix{URL }\fi
\providecommand{\eprint}[2][]{\url{#2}}

\bibitem[{{Chesneau} et~al.(2012){Chesneau}, {Lagadec}, {Otulakowska-Hypka},
  {Banerjee}, {Woodward}, {Harvey}, {Spang}, {Kervella}, {Millour}, {Nardetto},
  {Ashok}, {Barlow}, {Bode}, {Evans}, {Lynch}, {O'Brien}, {Rudy}, \&
  {Russell}}]{2012A&A...545A..63C}
{Chesneau}, O., {Lagadec}, E., {Otulakowska-Hypka}, M., {Banerjee}, D.~P.~K.,
  {Woodward}, C.~E., {Harvey}, E., {Spang}, A., {Kervella}, P., {Millour}, F.,
  {Nardetto}, N., {Ashok}, N.~M., {Barlow}, M.~J., {Bode}, M., {Evans}, A.,
  {Lynch}, D.~K., {O'Brien}, T.~J., {Rudy}, R.~J., \& {Russell}, R.~W. 2012,
  \aap, 545, A63

\bibitem[{{Das} et~al.(2008){Das}, {Banerjee}, {Ashok}, \&
  {Chesneau}}]{2008MNRAS.391.1874D}
{Das}, R.~K., {Banerjee}, D.~P.~K., {Ashok}, N.~M., \& {Chesneau}, O. 2008,
  \mnras, 391, 1874

\bibitem[{{Hachisu} \& {Kato}(2004)}]{2004ApJ...612L..57H}
{Hachisu}, I., \& {Kato}, M. 2004, \apjl, 612, L57

\bibitem[{{Hachisu} et~al.(2008){Hachisu}, {Kato}, \&
  {Cassatella}}]{2008ApJ...687.1236H}
{Hachisu}, I., {Kato}, M., \& {Cassatella}, A. 2008, \apj, 687, 1236

\bibitem[{{Iijima}(2006)}]{2006A&A...451..563I}
{Iijima}, T. 2006, \aap, 451, 563

\bibitem[{{Naito} et~al.(2012){Naito}, {Mizoguchi}, {Arai}, {Tajitsu},
  {Narusawa}, {Yamanaka}, {Fujii}, {Iijima}, {Kinugasa}, {Kurita}, {Nagayama},
  {Yamaoka}, \& {Sadakane}}]{2012A&A...543A..86N}
{Naito}, H., {Mizoguchi}, S., {Arai}, A., {Tajitsu}, A., {Narusawa}, S.,
  {Yamanaka}, M., {Fujii}, M., {Iijima}, T., {Kinugasa}, K., {Kurita}, M.,
  {Nagayama}, T., {Yamaoka}, H., \& {Sadakane}, K. 2012, \aap, 543, A86

\bibitem[{{Naito} \& {Narusawa}(2007)}]{2007IAUC.8803....2N}
{Naito}, H., \& {Narusawa}, S. 2007, IAU Circ., 8803, 2

\bibitem[{{Naito} et~al.(2013){Naito}, {Tajitsu}, {Arai}, \&
  {Sadakane}}]{2013PASJ...65...37N}
{Naito}, H., {Tajitsu}, A., {Arai}, A., \& {Sadakane}, K. 2013, \pasj, 65, 37

\bibitem[{{Sadakane} et~al.(2010){Sadakane}, {Tajitsu}, {Mizoguchi}, {Arai}, \&
  {Naito}}]{2010PASJ...62L...5S}
{Sadakane}, K., {Tajitsu}, A., {Mizoguchi}, S., {Arai}, A., \& {Naito}, H.
  2010, \pasj, 62, L5

\bibitem[{{Yamaoka} et~al.(2007){Yamaoka}, {Nakamura}, {Nakano}, {Sakurai}, \&
  {Kadota}}]{2007IAUC.8803....1Y}
{Yamaoka}, H., {Nakamura}, Y., {Nakano}, S., {Sakurai}, Y., \& {Kadota}, K.
  2007, IAU Circ., 8803, 1

\end{thebibliography}

\end{document}